# Enhanced Performance and Stability of Perovskite Solar Cells with Ag-Cu-Zn Alloy Electrodes

*Keshav Kumar Sharma, Ashutosh Ujjwal, Rohit Saini, and Ramesh Karuppannan\**

Department of Physics, Indian Institute of Science, Bangalore 560012, Karnataka, India
\*E-mail: kramesh@iisc.ac.in

**Abstract:** Though the common metal electrode-based perovskite solar cells have achieved a power conversion efficiency of >25%, they also play a crucial role in accelerating the degradation of the cells. In this study, we investigated phase transition engineering in Ag electrodes via Cu and Zn alloying, transforming from a cubic to a tetragonal phase. These alloyed electrodes are then thermally deposited as back electrodes in perovskite solar cells. We conducted a comprehensive analysis of the pure Ag and Ag-Cu-Zn alloys deposited atop a hole-transport layer for use in $Cs_{0.05}(FA_{0.83}MA_{0.17})_{0.95}Pb(I_{0.83}Br_{0.17})_3$-based solar cells. Our findings reveal that solar cells developed with pure Ag electrodes demonstrate a power conversion efficiency (PCE) of 18.71%, characterized by a fill factor (FF) of 74.8%, an open-circuit voltage ($V_{OC}$) of 1.08 V, and a short-circuit current density ($J_{SC}$) of 23.17 mA/cm². Conversely, solar cells fabricated with optimized $Ag_{0.875}Cu_{0.120}Zn_{0.005}$ electrodes exhibit enhanced performance metrics, with an FF of 72.5%, $V_{OC}$ of 1.12 V, and $J_{SC}$ of 23.39 mA/cm², culminating in an elevated PCE of 19.02%. Moreover, this electrode demonstrates remarkable durability, sustaining operational integrity for 460 hours for the PSCs stored in $N_2$ glove box, in contrast to the 320 hours for cells with Ag electrodes. The Ag-Cu-Zn alloys exhibited high resistance to corrosion and good adhesion on the hole-transport material layer compared to a layer of Ag. These advancements may lead to the realization of cost-effective, durable, and efficient solar energy conversion systems.

## 1. Introduction

Perovskite solar cells (PSCs) have recently been demonstrated as potential candidates for the next generation of green and renewable energy conversion due to their low material and manufacturing cost, scalable fabrication capability, and very high efficiency achieved[1,2]. The long-term stability of perovskite materials is still a major challenge for commercialization[3,4]. Factors such as heat, oxygen and moisture lead to degradation of the perovskite materials and have garnered significant attention from researchers due to their significant impact on the stability and longevity of perovskite-based devices[5,6]. Additionally, there is growing attention towards understanding the mechanisms behind the reduced stability of perovskite solar cells, particularly due to ion migration between the electrode and the perovskite layer[7,8].

The metals with suitable work function and conductivity have been chosen as electrodes in PSCs. The metallic layers of Al, Au, Cu and Ag have been reported to be used regularly for back-contact electrodes in PSCs. Properties of different materials, interfaces, and morphology of the electrodes influence the overall device performance. Al is not a preferred material for PSCs due to its susceptibility to corrosion and oxidation[9,10]. Al may also react with degradation products of the perovskite layer producing organo-aluminum compounds or alternatively anion radicals. These compounds can easily react with any proton donor present (e.g. a trace of water or with oxygen)[11]. Ag electrode is broadly used in the PSCs, but its reactive and corrosive nature with perovskite consistently induces the critical stability issue[12,13]. As well, Ag atoms diffuse into the perovskite layer, they can accelerate the reaction with iodine species, and induce the deep level defects in the perovskite layer[14,15]. Cu is a more appropriate electrode material for higher stability of PSCs than conventional Al and Ag electrodes. Cu has been observed to diffuse into and react with the perovskite layer, and the infrared and ultraviolet contents in the sunlight accelerate the photo-oxidation and chemical reaction between Cu and the perovskite layer[10,16]. Therefore, Au has been considered a logical choice of the electrode as it inherently provides resistance to environmental oxidation and appears also to be resistant to metal–halide formation, however diffusion of Au into the perovskite active layer can induce deep traps, leading to a degradation in performance[10,17,18]. Au is a relatively expensive material, which can increase the overall cost of PSCs, making them less economically viable for large-scale production[19]. In this study, we present the incorporation of Ag-Cu-Zn alloy compositions, encompassing variants of $Ag_{0.875}Cu_{0.125}Zn_{0.000}$, $Ag_{0.875}Cu_{0.120}Zn_{0.005}$, and $Ag_{0.875}Cu_{0.115}Zn_{0.010}$, as novel electrode materials. These alloys show excellent thermal and electrical conductivity and good reflectivity of light and present a cost-effective alternative to other precious electrode metals. Ag-Cu-Zn alloy based back electrodes are less corrosive and less diffusive relative to Ag electrodes and reduce the device degradation in the unencapsulated PSCs under ambient conditions.

In this work, we report the innovative demonstration of thermal evaporation processing of Ag-Cu-Zn alloy as a novel back electrode material for high-efficiency and long-term stable $Cs_{0.05}(FA_{0.83}MA_{0.17})_{0.95}Pb(I_{0.83}Br_{0.17})_3$-based PSCs. The PSCs fabricated with Ag and Ag-Cu-Zn electrodes are stored in the dark under ambient air conditions. Notably, PSCs with Ag electrodes exhibit degradation with time, which is attributed to Ag corrosion, as evidenced by the formation of silver iodide (AgI) confirmed through XRD and XPS analysis. In contrast, alloying Ag with Cu and Zn significantly reduce the corrosion rate of the electrodes, minimizes the diffusion of Ag atoms into the perovskite layer, and consequently enhances device stability.

## 2. Experimental Section

**Figure 1** illustrates the schematic representation of the fabrication process of the Ag-Cu-Zn alloy back electrodes via the melt quenching technique followed by thermal evaporation. To prepare the Ag-Cu-Zn alloys, the appropriate amounts of Ag, Cu, and Zn powders were grinded, as described in Table S1 provided in the supporting information and then transferred into cleaned and flat quartz ampoules. The ampoules were subsequently sealed under a vacuum of $10^{-6}$ mbar and loaded into a resistive furnace. The furnace temperature was initially ramped to 350 °C and maintained for 3 hours, followed by a gradual increase to 700 °C. After a 3-hour isothermal hold at 700 °C, the temperature was further elevated to 1050 °C. To ensure sample homogeneity, the ampoules were continuously rocked throughout the process. Following a 12-hour dwell at 1050 °C, the temperature was reduced to 800 °C, and the melt was rapidly quenched in an ice-water bath. Post-quenching, the Ag-Cu-Zn alloys were annealed at 250 °C to relieve stress and improve homogenization.

To fabricate the PSCs, the precursor solutions of $SnO_2$, $Cs_{0.05}(FA_{0.83}MA_{0.17})_{0.95}Pb(I_{0.83}Br_{0.17})_3$ (abbreviated as CsFAMA), and spiro-OMeTAD were prepared for the deposition of the electron transport layer (ETL), active layer, and hole transport layer (HTL), respectively. The detailed procedures for preparing these precursor solutions are outlined in the Methods section of the supporting information. The step-by-step fabrication process utilized for the preparation of PSCs is illustrated in the schematic diagram presented in **Figure S1**.

After optimization, when silver alloying with copper and Zinc, the composition of $Ag_{0.875}Cu_{0.120}Zn_{0.005}$ emerged as the most efficacious electrode material for PSCs. This alloy and its corresponding thin film/device are denoted as ACZ0.5/target, while fine silver serves as Ag/control.

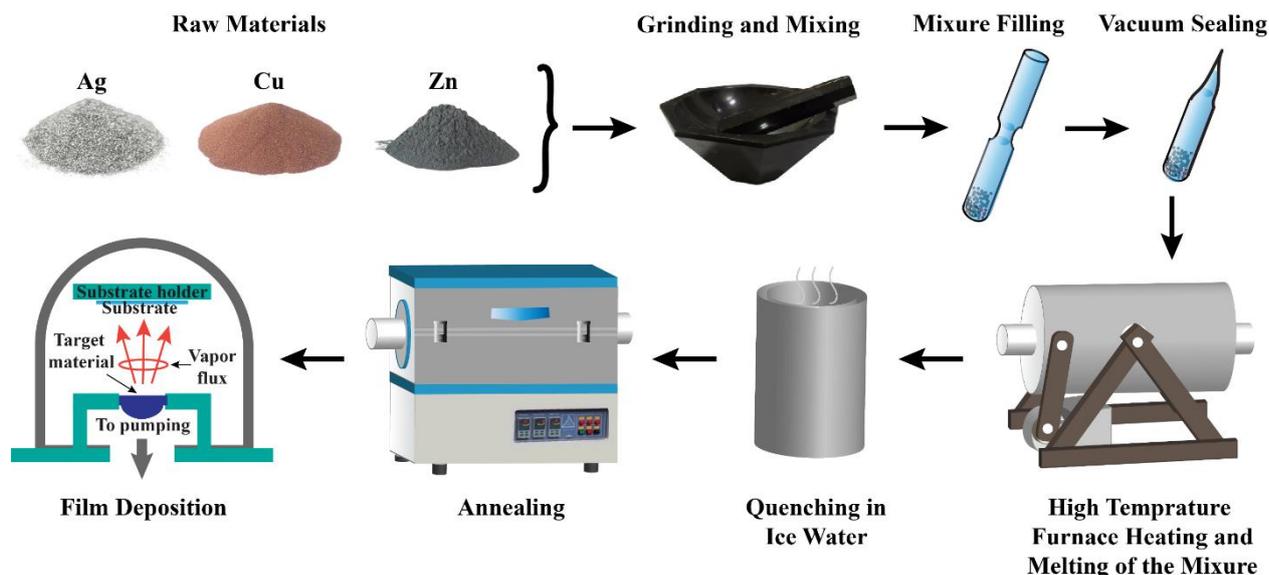

**Figure 1.** Schematic diagram of the fabrication of Ag-Cu-Zn alloy back electrodes.

## 3. Results and Discussion

The X-ray diffraction (XRD) patterns of both Ag and ACZ0.5 thin films were subjected to analysis using the pseudo-Voigt function through the Rietveld refinement method. The Rietveld refinement results for both thin films are depicted in **Figure 2a,b**. The results of the refinement for fine silver confirm a face-centered cubic structure with a space group *Fm-3m* (No. 225) and unit cell parameters a=b=c=4.09 Å. In contrast, ACZ0.5 exhibits a crystalline structure with a tetragonal symmetry, characterized by a space group *P4/mmm* (No. 123) and lattice parameters a=b=4.074 Å and c=8.148 Å, respectively. The observed peaks align well with the standard files and reported theoretical results[20]. The refinement outcomes, including reliability factors $R_P$, $R_{wP}$, Goodness of Fitting (GOF), R Bragg, $\chi^2$, $R_F$, and lattice parameters for both samples, are summarized and tabulated in Table S2. Peak shifting and broadening can be seen in Ag thin films when alloyed with Cu and Zn (refer to **Figure S2**) and the corresponding crystal structures are illustrated in **Figure S3**.

The electronic structure and surface chemical compositions of Ag and the ACZ0.5 alloy were characterized using X-ray photoelectron spectroscopy (XPS). The XPS full survey of ACZ0.5 confirms the presence of Ag, Cu and Zn elements (refer to **Figure S4**). The high-resolution XPS spectrum of Ag 3d illustrated in **Figure 2c** shows that ACZ0.5 comprises two distinct peaks at approximately 367 and 375 eV, corresponding to Ag $3d_{5/2}$ and Ag $3d_{3/2}$, respectively. The separation between these peaks is calculated to be 8.0 eV, confirming the $Ag^0$ state[21]. The high-resolution XPS spectrum of Cu 2p depicted in **Figure 2d** shows that ACZ0.5 contains two distinct peaks at approximately 933 and 953 eV, corresponding to Cu $2p_{3/2}$ and Cu $2p_{1/2}$, respectively. The separation between the Cu $2p_{3/2}$ and Cu $2p_{1/2}$ peak is calculated to be 20.0 eV, which confirm the $Cu^0$ state[22,23]. The high-resolution XPS of Zn 2p exhibits double peaks located at the binding energy of 1021 and 1045 eV with a spin–orbit splitting energy of 24 eV, corresponding to the Zn $2p_{1/2}$ and Zn $2p_{3/2}$, respectively, as shown in **Figure 2e**. The Zn $2p_{3/2}$ peak can be deconvoluted into two peaks centered at 1021 and 1022 eV, which can be assigned to $Zn^0$ and $Zn^{2+}$, respectively. Similarly, the Zn $2p_{1/2}$ peak can be deconvoluted into two peaks centered at 1045 eV and 1046 eV[24,25]. The O 1s spectrum of ACZ0.5 alloy can be deconvoluted into two peaks, as illustrated in **Figure 2f**, corresponding to the binding energy of Zn-O (530 eV) and Zn-OH (532.5 eV) [25,26]. The formation of ZnO on the ACZ0.5 surface enhances the corrosion resistance of the electrode. Raman spectra of Ag and its alloys also confirm the formation of ZnO on the surface (refer to **Figure S5**). The atomic ratio of Ag/Cu was calculated to be 6.9, which is close to the stoichiometric value, while the atomic concentration of Zn decreases with depth.

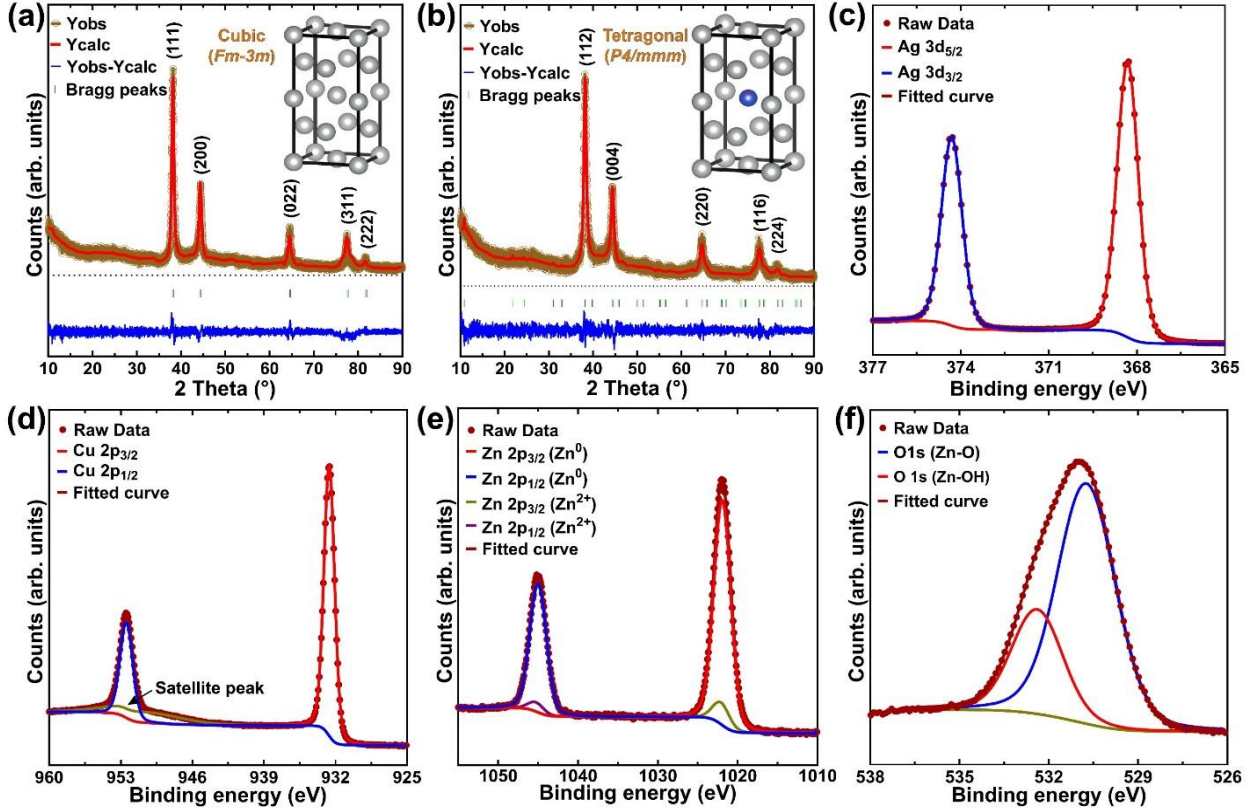

**Figure 2.** The Rietveld refinement fits of the XRD patterns of (a) the fine silver, cubic (*Fm-3m*), and (b) the ACZ0.5 alloy, tetragonal (*P4/mmm*) thin films. XPS spectra of ACZ0.5 alloy (c) Ag 3d, (d) Cu 2p, (e) Zn 2p, and (f) O 1s.

The Kelvin Probe Force Microscopy (KPFM) technique was utilized for the localized determination of surface potential and work function on the nanoscale. In an ideal scenario, KPFM quantifies the Contact Potential Difference (CPD) between the metallic Atomic Force Microscopy (AFM) tip and the sample, expressed by the relation[27]:

$$V_{CPD} = (\Phi_{tip} - \Phi_{sample}) / e^-  \quad (1)$$

Here, $\Phi_{tip}$ and $\Phi_{sample}$ denote the work functions of the sample and the tip, respectively, and e- represents the elementary charge. **Figures 3a-f** show the KPFM images, CPD profiles and corresponding morphological mapping, including height profiles, for Highly Oriented Pyrolytic Graphite (HOPG), Ag, and ACZ0.5 alloy. To illustrate the work function, a Nanosensors PtIr-PPP-EFM probe, resonating at ca. 65.7 kHz, was employed to obtain CPD and corresponding morphological mapping over each sample. HOPG was employed for the calibration of the tip work function. The CPD of HOPG was determined to be 299 mV, as illustrated in **Figure 3a**, and the work function of HOPG was previously reported to be ~4.6 eV under environmental conditions, as cited in the preceding literature[27–29]. The absolute surface work function of the sample was calculated with the following equation[27]:

$$\Phi_{sample} = 4.6 \text{ eV} + e^-(V_{CPDHOPG} - V_{CPDsample})  \quad (2)$$

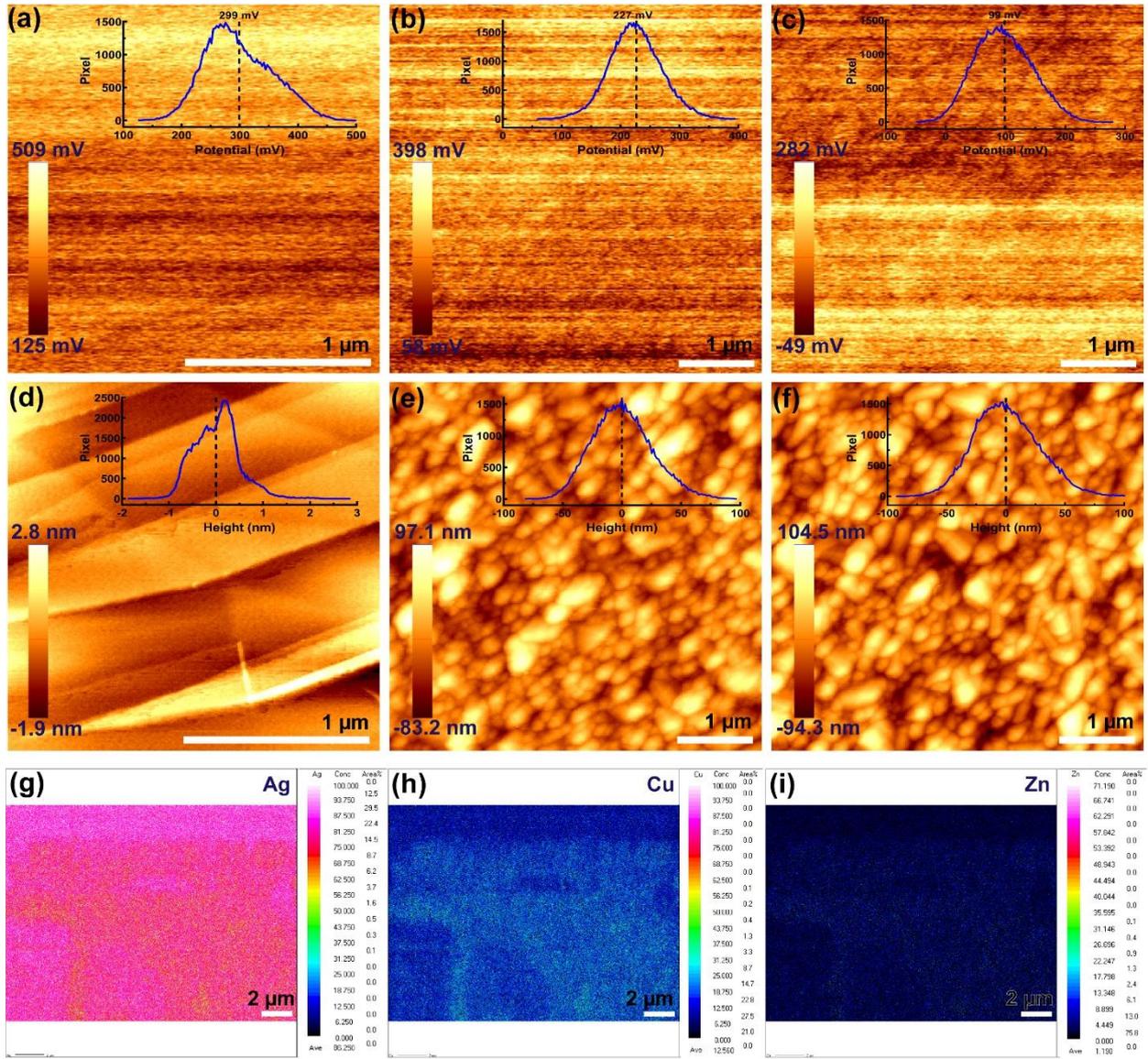

**Figure 3.** KPFM images and CPD profiles (a) HOPG, (b) Ag, and (c) ACZ0.5 alloy. AFM image and height profiles of (d) HOPG, (e) Ag, and (f) ACZ0.5 alloy. EPMA analysis of ACZ1.0 alloy sample on glass substrate: WDS mapping images showing elemental distribution: (g) Ag, (h) Cu, and (i) Zn.

With the above information, the work function of the tip was calculated to be $\varPhi_{tip}$ = 4.899 eV. This value is consistent with previous reports[27]. The CPD values for Ag and ACZ0.5 alloy were measured at 227 mV and 99 mV, as depicted in **Figure 3b,c**. Using $\varPhi_{tip}$ and CPD values, the work function values of fine silver and ACZ0.5 alloy were calculated to be $\varPhi_{Ag}$ = 4.67 eV and $\varPhi_{ACZ0.5}$ = 4.80 eV, respectively. These values of work function are consistent with UPS analysis (refer to **Figure S6**). Surface morphology images obtained through AFM for HOPG, Ag, and ACZ0.5 alloy are shown in **Figure 3d-f**, respectively. Both Ag and ACZ0.5 alloy thin films exhibit similar morphologies, with roughness measurements of 1.3 nm and 1.4 nm, respectively (refer to **Figure S7**).

Electron probe microanalysis (EPMA) was conducted to analyze the elemental distribution in the alloy sample. The EPMA mapping results of ACZ1.0 in at% are shown in Figure **3g-i**, and corresponding area of map back-scattered electron (BSE) and secondary electron (SE) are shown in **Figure S8**. EPMA mapping indicates that the distributions of Ag, Cu, and Zn are uniform. The overall atomic fractions of Ag, Cu and Zn in the ACZ1.0 alloy were measured to be 0.8625, 0.1256, and 0.0119, respectively, which is consistent with the theoretical composition of the ACZ1.0 alloy. These findings confirm that the alloy predominantly consists of a single phase.

To study of the moisture-induced degradation of electrodes with time, devices with architecture FTO/SnO$_2$/CsFAMA/spiro-OMeTAD/electrode were fabricated as described in the Methods in the supporting information. For this analysis, Ag and ACZ0.5 electrodes with precisely controlled thicknesses of approximately 8 nm (for XPS analysis) and 100 nm (for XRD analysis) were deposited via thermal evaporation. We measured XRD and XPS for both the freshly prepared samples and those subjected to aging for specific durations at a relative humidity of 40% in the ambient air. The measurements were performed from the top electrode side to discern any alterations in the structural and chemical characteristics of the control and target devices, as illustrated in **Figure 4a**.

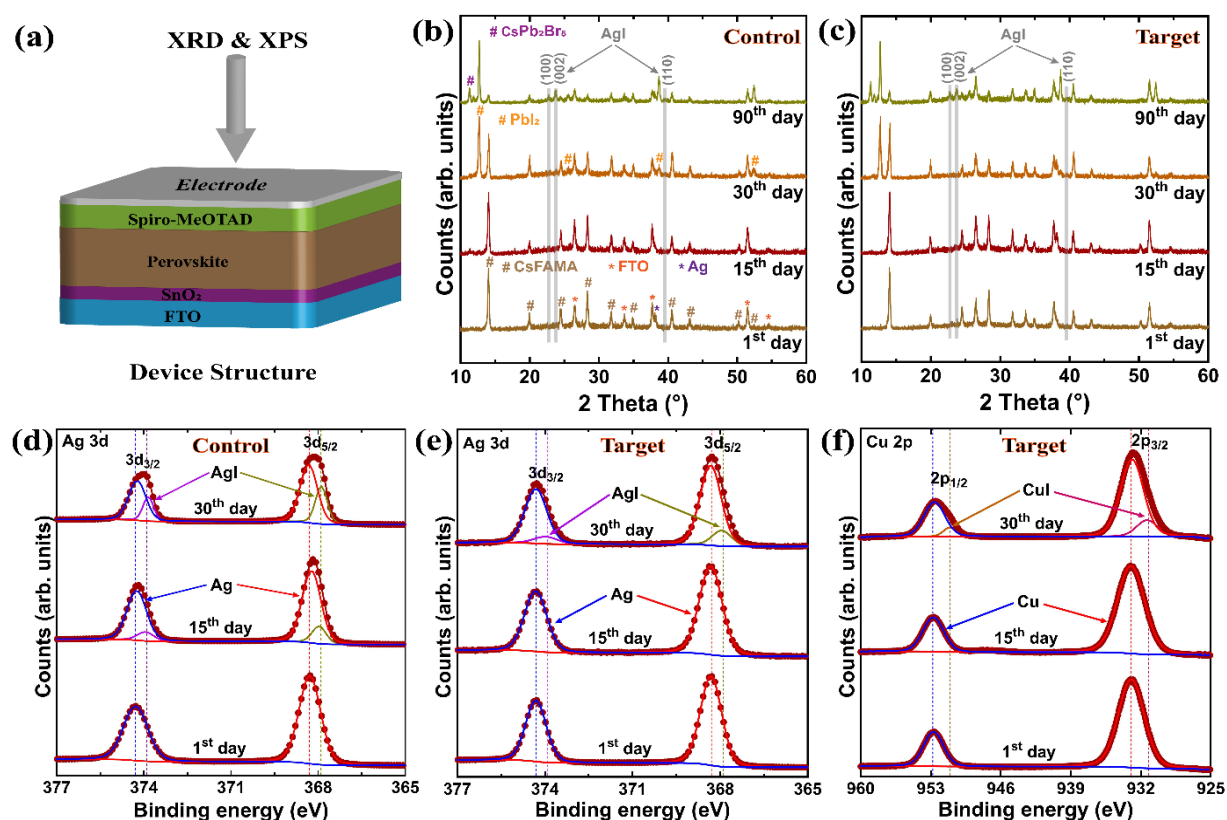

**Figure 4.** (a) Schematic structure of the device with Ag/ACZ0.5 electrode, (b) XRD patterns of the device with Ag (control) electrode up to 90 days, (c) XRD patterns of the device with ACZ0.05 (target) electrode up to 90 days, (d) Ag 3d XPS spectrum of the device with Ag (control) electrode up to 30 days, (e) Ag 3d XPS spectrum of the device with ACZ0.5 (target) electrode up to 30 days, and (f) Cu 2p spectrum of the device with ACZ0.5 (target) electrode up to 30 days.

XRD patterns exhibit variation as a function of time (see **Figure 4b,c**). Initially, it consisted of combining XRD patterns from FTO, CsFAMA, Ag and ACZ0.5 for both control and target devices (refer to **Figure S9a,c**, and Figure 2a,b). As time progressed, notable changes in the XRD patterns were observed, suggesting evolving structural characteristics within the CsFAMA film and electrodes. After 30 days, the control device stored in ambient air showed XRD peaks at 12.65, 25.55, 38.67 and 52.48°, which correspond to the crystallographic planes of PbI$_2$, specifically (001), (002), (003) and (004), respectively (refer to **Figure S9d**).

Additionally, a peak at 23.80° was observed, which was indexed to the (002) crystallographic plane of AgI. Notably, this peak persisted with two new peaks at 22.68 and 39.46° in the XRD pattern of the control device even after an extended storage period of 90 days in ambient air[12,21,30–32]. These peaks can be assigned to the (100) and (110) crystallographic plane of AgI, respectively. Furthermore, one additional peak appeared at 11.32° after storage period of 90 days, which corresponds to the (002) crystallographic planes of $CsPb_2Br_5$[33]. Similarly, the XRD analysis of the target device, stored under the same conditions, revealed the presence of identical XRD peaks with storage period. Interestingly, no AgI peaks were seen after 30 days of storage. However, after 90 days, all three AgI peaks appeared. This observation suggests that the ACZ0.5 alloy electrode exhibits enhanced stability compared to the Ag electrode.

On the other hand, the fresh devices (FTO/SnO$_2$/CsFAMA/spiro-OMeTAD/electrode) did not exhibit any discernible peaks corresponding to AgI (see Figure 4b,c). Both control and target devices show notable peaks at 38.12 and 38.19°, respectively, attributed to Ag (111) and ACZ0.5 (112) crystallographic plane[12,32]. These peaks were observed continuously in both devices stored for a period of 90 days at a relative humidity of 40% in ambient air. This observation confirms that Ag was not converted completely to AgI after a storage period of 90 days.

To investigate the degradation mechanism of electrodes over time, XPS analysis was conducted on both devices for a duration of 30 days, as depicted in **Figure 4d,e,f**. Relatively a large amount of iodine and bromine were detected after storage period of 15 days and 30 days, respectively, in both control and target device (refer to **Figure S12**). This phenomenon is likely attributed to the presence of pinholes in the spiro-OMeTAD films, enabling the detection of the underlying exposed perovskite layer. The Ag 3d XPS spectrum of fresh control device revealed two distinct and well-resolved peaks at 368.3 eV and 374.3 eV for Ag $3d_{5/2}$ and $3d_{3/2}$, respectively (see **Figure 4d**). No noticeable peak shifting in Ag 3d XPS spectrum of fresh target device was observed, as shown in **Figure 4e**. Additionally, the XPS spectrum of Cu 2p exhibited two prominent peaks at 952.8 eV and 933.0 eV for Cu $2p_{3/2}$ and $2p_{1/2}$, respectively, with a separation of 19.8 eV between them (see **Figure 4f**). After storage period of 15 days, a notable observation was made in the control device. The Ag 3d peaks clearly shift from a metallic state to a lower binding energy. This shift is characteristic of the formation of AgI, as illustrated in Figure 4d. Interestingly, a similar shifting phenomenon in the target device was observed, but it occurred after storage period of 30 days rather than 15 days, as depicted in Figure 4e. Further, a metallic to lower binding energy shifting in Cu 2p peaks was observed after a storage period of 30 days, which is the characteristics of CuI, as shown in Figure 4f. The formation of CuI at the spiro-OMeTAD/electrode interface facilitates hole transport[34,35]. These findings also confirm that ACZ0.5 electrode is highly stable compared to Ag electrode.

XRD and XPS analysis reveals that device degradation occurs in a two-step process. In the first step, the CsFAMA layer undergoes degradation upon reacting with moisture, which penetrates through the pinholes present in the spiro-OMeTAD layer. During the second step, the degraded constituents of CsFAMA, specifically $PbI_2$, migrate through the spiro-OMeTAD layer and subsequently react with Ag, leading to the formation of AgI through a diffusion process. The conceptual model of degradation of Ag electrode deposited on the FTO/SnO$_2$/CsFAMA/spiro-OMeTAD is illustrated in **Figure S13**. Furthermore, scanning electron microscopy (SEM) images of the back electrodes of both the control and target devices, collected after 30 days, show that the Ag electrode degrades more rapidly compared to the ACZ0.5 electrode (refer to **Figure S14**). This confirms that the ACZ0.5 electrode is more stable than the Ag electrode.

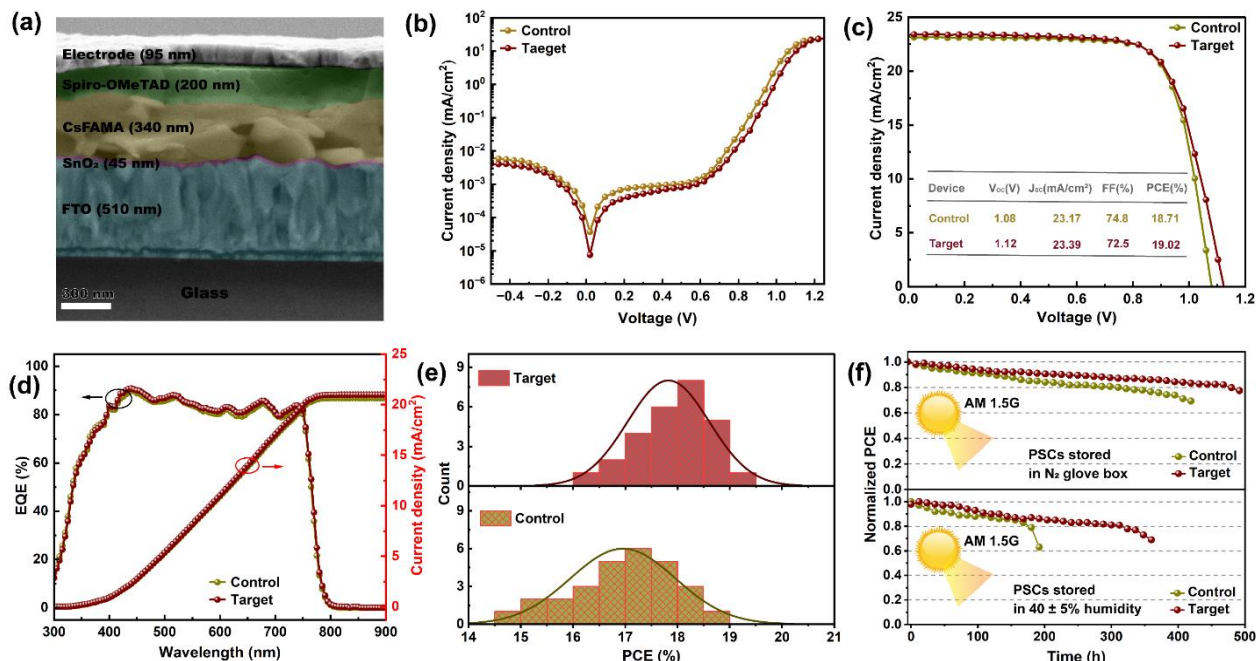

**Figure 5.** (a) Cross-sectional SEM image of PSCs, (b) Dark *J–V* characteristics of the control and target devices, (c) illuminated *J–V* measurement of control and target devices with reverse scan, (d) EQE and integrated current density curves, (e) the statistical distribution of the PCE and (f) stability measurements of control and target devices in the environment of 40 ± 5% humidity (bottom) and which stored in $N_2$ glove box (top).

We fabricated planar *n–i–p* PSCs with configuration FTO/$SnO_2$/CsFAMA/spiro-OMeTAD/electrode. **Figure 5a** shows a cross-sectional FE-SEM image of the PSC prepared by the solvent engineering method with CsFAMA as an active layer, $SnO_2$ as an ETL and spiro-OMeTAD as an HTL. The PSC consisted of a uniform layer of $SnO_2$ (~45 nm in thickness), a uniformly deposited CsFAMA layer (~340 nm in thickness) and a uniform layer of spiro-OMeTAD (~200 nm in thickness). We prepared PSCs with Ag and ACZ0.5 as a back electrode (~95 nm thick) for the control and target device, respectively. SEM image, absorption spectra, steady-state photoluminescence spectra, and time-resolved photoluminescence profiles of CsFAMA thin film are shown in **Figure S11**. The energy band diagram of the PSCs is shown in **Figure S15**[36,37].

Dark current-voltage characteristics, as illustrated in **Figure 5b**, were used to investigate the effect of the spiro-OMeTAD/electrode interface on the recombination nature in the device. The target device shows a lower leakage current ($6.3 \times 10^{-6}$ mA/cm$^2$) than the Ag electrode-based device ($3.3 \times 10^{-5}$ mA/cm$^2$) indicating the degraded recombination at the spiro-OMeTAD/Ag interface[38,39]. The reduction in leakage current observed in the target device can be attributed to improved chemical stability, better adhesion, and reduced diffusion of silver ions into the perovskite layer. In **Figure 5c**, the photocurrent density-voltage (*J–V*) characteristics of the PSCs fabricated with the ACZ0.5 electrode as target device are compared with those of the control device prepared with the Ag electrode. As shown in the *J–V* curves, the target device exhibited an open-circuit voltage ($V_{OC}$) of 1.12 V, a short-circuit current density ($J_{SC}$) of 23.39 mA/cm$^2$, and a fill factor (FF) of 72.5%, giving an maximum PCE of 19.02%, whereas the control device showed an overall PCE of 18.71% with a $V_{OC}$ of 1.08 V, a $J_{SC}$ of 23.17 mA/cm$^2$, and a FF of 74.8%. The superior performance of the target device is mainly attributed to a higher $V_{OC}$, which is associated with beneficial effects such as electronic structure modification and chemical stability of the ACZ0.5 electrode. The lower FF of the target device may be attributed to the higher sheet resistance (0.012 ohm/sq for Ag and 0.425 ohm/sq for ACZ0.5) of the ACZ0.5 electrode compared to the Ag electrode. The external quantum efficiency (EQE) of the control and target devices was tested, as shown in **Figure 5d**. The results reveal that the target device shows a good optical response with a higher EQE and an integrated current density of 21.78 mA/cm$^2$, while the control device exhibits an integrated current density of 21.30 mA/cm$^2$. The statistical distribution diagram of the PCE of the PSCs is shown in **Figure 5e** and the *J*–*V* data of 28 devices is shown in Table S3, and the PCE results exhibit that the PSCs with ACZ0.5 electrodes show good reproducibility and better photovoltaic performance. Time-stability plots of the PSCs are shown in **Figure 5f**. The control device, stored in the environment of 40 ± 5% humidity loses 20% of its efficiency after 180 hours, whereas the target device takes 310 hours to experience the same level of efficiency loss. The PCE of the control device, stored in the N$_2$ glove box, decreases to 80% of its initial value after 320 hours, while the PCE of the target device retains over 80% of its initial value after 460 h. These findings indicate that the target device exhibits superior stability, particularly in terms of moisture stability.

## 4. Conclusions

The ACZ0.5 alloy-based back electrode demonstrates remarkable potential for use in PSCs due to its excellent chemical stability, good reflectivity of light, and high adhesion on the hole transport layer. Our findings demonstrate that PSCs incorporating the ACZ0.5 alloy electrode exhibit significantly enhanced durability compared to those utilizing Ag metal electrodes, particularly in unencapsulated devices. Additionally, these cells achieve high efficiency, and a larger open-circuit voltage than traditional Ag electrodes. These attributes make the ACZ0.5 alloy-based electrode a promising candidate for photovoltaic applications, paving the way for more reliable, cost-effective and efficient solar energy solutions. Further research and development could enhance its performance and broaden its applicability in the renewable energy sector.

**Supporting Information**.
Supporting Information is available with this paper.

**Corresponding Author**
**Ramesh Karuppannan** – Department of Physics, Indian Institute of Science (IISc), Bangalore 560012, Karnataka, India; https://orcid.org/0000-0002-8304-6500; Email: kramesh@iisc.ac.in

**Authors**
**Keshav Kumar Sharma** – Department of Physics, Indian Institute of Science (IISc), Bangalore 560012, Karnataka, India; https://orcid.org/0000-0002-6753-2269
**Ashutosh Ujjwal** – Department of Physics, Indian Institute of Science (IISc), Bangalore 560012, Karnataka, India; https://orcid.org/0009-0001-7786-4179
**Rohit Saini** – Department of Physics, Indian Institute of Science (IISc), Bangalore 560012, Karnataka, India; https://orcid.org/0009-0000-7004-0401

**Author Contributions**
K. K. Sharma designed the study, conducted experiments, analyzed data, and wrote the manuscript. A. Ujjwal and R. Saini contributed to material characterization. K. Ramesh conceptualized the study, provided guidance throughout the research process, by conducting thorough reviews. All authors contributed to reviewing the manuscript.

**Acknowledgments**
We acknowledge support from CeNSE facilities funded by MHRD, MeitY and DST Nano Mission. We thank Indian Science Technology and Engineering facilities Map (I-STEM), a Program supported by Office of the Principal Scientific Adviser to the Govt. of India, for

# Supporting Information

# Enhanced Performance and Stability of Perovskite Solar Cells with Ag-Cu-Zn Alloy Electrodes

*Keshav Kumar Sharma, Ashutosh Ujjwal, Rohit Saini, and Ramesh Karuppannan\**

Department of Physics, Indian Institute of Science, Bangalore, Karnataka, India 560012
*E-mail: kramesh@iisc.ac.in

## 1. Methods
### 1.1 Materials and solvents

Zinc powder (Zn, 96%), lead iodide ($PbI_2$, 98%), and lead bromide ($PbBr_2$, 98%) were purchased from TCI chemicals. Methylammonium bromide (MABr, 99.99%), formamidinium iodide (FAI, 99.99%) were purchased from GreatCell Solar. Cesium iodide (CsI, 99.999%) and Tin(IV) chloride pentahydrate ($SnCl_4.5H_2O$, 98%), 2,2',7,7'-tetrakis [N, N-di(4-methoxyphenyl) amino]-9,9'-spirobifluorene (Spiro-OMeTAD, 99%) were purchased from Sigma-Aldrich. N, N-Dimethyl formamide (DMF, 99.8%), dimethyl sulfoxide (DMSO, 99.9%), acetonitrile (ACN), 4-tert-butylpyridine (t-BP), Ethanol ($C_2H_5OH$), hydrochloric acid (HCl, 37%) and Acetylacetone (acacH, ≥99%) were also purchased from Sigma-Aldrich without further purification. Chlorobenzene (CB, 98.0%) and lithium bis (trifluoromethanesulfonyl) imide (Li-TFSI, 98.0%) were also purchased from TCI chemicals. Fluorine-doped tin oxide (FTO, 7 Ω/sq.) and Hellmanex III were also purchased from Sigma-Aldrich. Fine silver (Ag, 99.9%), Copper (99.9%) and Zinc (98%) were purchased from commercial sources.

### 1.2 $SnO_2$ precursor solution

The precursor solution was prepared by dissolving 31.5 mg of $SnCl_4.5H_2O$ in 1 mL absolute ethanol. 20.5 μL of acacH was added as chelating at room temperature to the solution to yield [acacH]/[Sn] > 2. In the presence of an excess of acacH, the hydrolytic stability of the tin-acac-chelate complex increases, preventing the progress of further condensation reaction. The nominal hydrolysis ratio of alcoholic solution prepared from $SnCl_4.5H_2O$ ($h = [H_2O]/[Sn] = 5$) was adjusted by dropwise addition of 162 μL of DI water to yield $h = [H_2O]/[Sn] = 105$. The solution was stirred at 70 °C for 2 hrs, giving rise to transparent and stable colloidal solution[1,2].

### 1.3 $Cs_{0.05}(MA_{0.17}FA_{0.83})_{0.95}Pb(I_{0.83}Br_{0.17})_3$ precursor solution

A solution was prepared by dissolving 507.1 mg of $PbI_2$ and 73.4 mg of $PbBr_2$ in 1 mL of anhydrous DMF and DMSO in a 4:1 (v/v) ratio, stirring at 100 °C for 30 minutes. After cooling, 172 mg of FAI, 22.4 mg of MABr, and 53 μL of a CsI solution (1.5 M in DMSO) were added to the inorganic solution to create the perovskite precursor solution. The precursor solution was then stirred at room temperature for 6 hrs[3].

### 1.4 Spiro-OMeTAD precursor solution

A solution of spiro-OMeTAD precursor was prepared by dissolving 72.3 mg of spiro-OMeTAD, 17.5 μL of Li-TFSI solution (520 mg in 1 mL ACN), and 28.8 μL of t-BP in 1 mL of CB[3].

### 1.5 Ag-Cu-Zn alloy preparation

To prepare the back electrode, silver, copper, and zinc metals were utilized in the specified compositions, as detailed in the table below:

**Table S1.** Composition of alloys when Ag alloying with Cu and Zn:

| S. No. | Ag (wt%) | Cu (wt%) | Zn (wt%) | Alloy (atomic friction) | Symbol |
|---|---|---|---|---|---|
| 1 | 100 | 0 | 0 | Ag | Ag |
| 2 | 92.22 | 7.78 | 0 | $Ag_{0.875}Cu_{0.125}$ | ACZ0.0 |
| 3 | 92.22 | 7.45 | 0.33 | $Ag_{0.875}Cu_{0.120}Zn_{0.005}$ | ACZ0.5 |
| 4 | 92.22 | 7.14 | 0.64 | $Ag_{0.875}Cu_{0.115}Zn_{0.010}$ | ACZ1.0 |

### 1.6 Solar cell fabrication

Solar cells were fabricated on FTO-coated glasses, which underwent sequential ultrasonic cleaning with Hellmanex III, deionized water and IPA for 15 minutes each. Subsequently, the FTO glasses were subjected to UV ozone cleaning for 30 minutes before being used. A compact layer of $SnO_2$ was prepared by spin coating of $SnO_2$ precursor solution at 3000 RPM for 30 secs on FTO glass, followed by annealing at 145 °C for 1 hr. After cooling down, it was UV ozone treated for 30 minutes before being used. The perovskite precursor solution was applied on FTO/$SnO_2$ layer through a two-step spin-coating process. In the first step, the coating was performed at 1000 RPM for 10 secs with a ramping rate of 500 RPM/sec. Subsequently, in the second step, the spin-coating process was carried out at 6000 RPM for 20 secs with a ramping rate of 2000 RPM/sec. During the second step, 200 μL of CB was poured onto the spinning substrate 5 secs prior to the end of spinning program, followed by the annealing at 110 °C for 1 hr on the hotplate. 20 μL of spiro-OMeTAD solution was spin-coated upon the perovskite layer at 500 RPM for 3 secs, and 3000 RPM for 30 secs. Finally, the devices were completed by thermal evaporation of ~100 nm thick fine silver and silver-copper-zinc alloys. The deposition rate during this process was set at 1 to 5 Å/sec under high vacuum ($6.5\times10^{-6}$ mbar). The active area of the devices, amounting to 0.09 cm$^2$, was defined using an evaporation mask.

### 1.7 Solar cell characterization

The current-voltage (*J–V*) characteristics of the unencapsulated solar cells were measured using a Keithley 2400 instrument. The measurement involved a forward scan from -0.5 V to 1.5 V at a scanning rate of 50 mV/s, with a voltage step of 10 mV and a delay time of 50 ms. The measurements were conducted under ambient conditions and illuminated with AM1.5G light at an intensity of 100 mW/cm$^2$ from a solar simulator (Newport). The solar simulator was calibrated using a standard silicon solar cell device.

### 1.8 X-rays diffraction (XRD)

XRD measurements were conducted using a Rigaku SmartLab diffractometer that was equipped with a copper Kα anode. The diffractometer operated at a tube output voltage of 45 kV and a current of 30 mA. Single scans were performed within the angular range of 10° to 80° using Bragg's angle, and the measurements were carried out in parallel beam (PB) geometry.

### 1.9 Raman spectroscopy

Raman spectra were obtained using a LabRAM HR evolution Raman microscope. A 532 nm argon ion laser was used as an excitation source for spectroscopic measurements.

### 1.10 Scanning Electron Microscopy (SEM)

The surface and cross-sectional morphologies of perovskite films deposited on $SnO_2$/FTO substrates were characterized using field-emission scanning electron microscopy (FE-SEM, ZEISS Ultra55, Mono Carl Zeiss). The perovskite films were fabricated following the same protocols employed in solar cell production.

## 1.11 Atomic Force Microscopy (AFM)

The surface morphologies and localized work functions of back electrode films were characterized using AFM with a Park NX20 system in Kelvin Probe Force Microscopy (KPFM) mode. A PtIr-PPP-EFM probe was employed as the scanning probe microscopy (SPM) probe.

## 1.12 X-ray photoelectron spectroscopy (XPS)

Chemistry of electrode degradation was examined by Thermo Scientific XPS/UPS system using an Al K$\alpha$ ($\lambda$ = 0.83 nm, h$\upsilon$ = 1486.7 eV). X-ray source was operated at 23.5 W, and the data were analyzed using CasaXPS software.

## 1.13 Ultraviolet photoelectron spectroscopy (UPS)

The work functions of thin films were calculated by Thermo-Scientific XPS/UPS system in UPS mode. UV photons are produced using a gas discharge lamp, typically filled with helium. He-I line having the energy of 21.22 eV was used for the spectroscopic measurements.

## 1.14 UV-Visible-NIR spectroscopy

The absorbance, transmittance, and reflectance spectra of thin films were collected by PerkinElmer LAMBDA 1050+ UV-Vis-NIR spectrometers. Aluminum coated glass was used as a reference for reflectance.

## 2. Device Fabrication

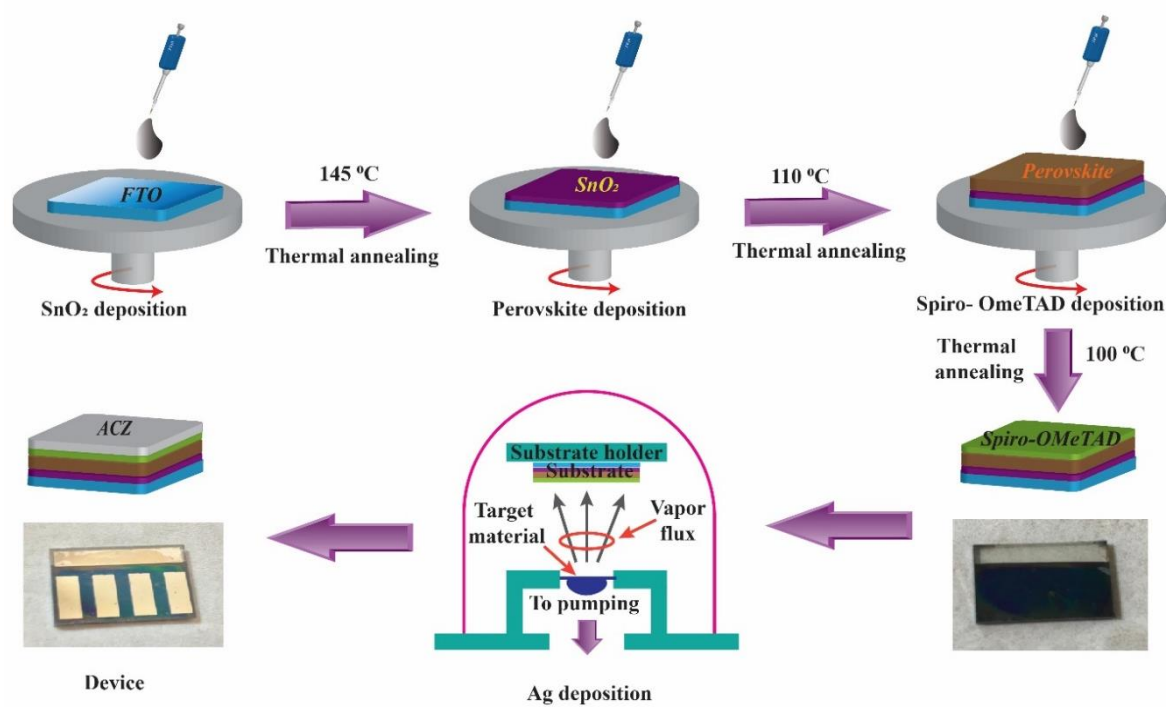

**Figure S1**. Schematic representation of the fabrication procedures for perovskite solar cells.

## 3. Characterization
### 3.1. Material characterization

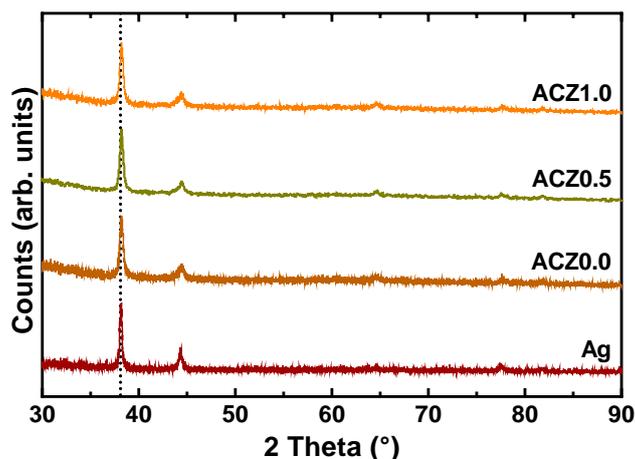

**Figure S2.** XRD spectra of (a) Ag, (b) ACZ0.0, (c) ACZ0.5, and (d) ACZ1.0.

Silver (Ag) initially exhibits a cubic phase with a space group *Fm-3m* (No. 225). When one silver atom out of eight is replaced by copper (Cu), the structure transitions to a tetragonal phase with a space group *P4/mmm* (No. 123). The addition of 0.5 to 1.0 % zinc (Zn) in place of Cu maintains the tetragonal phase. The crystal structures of all compositions are illustrated in **Figure S1**.

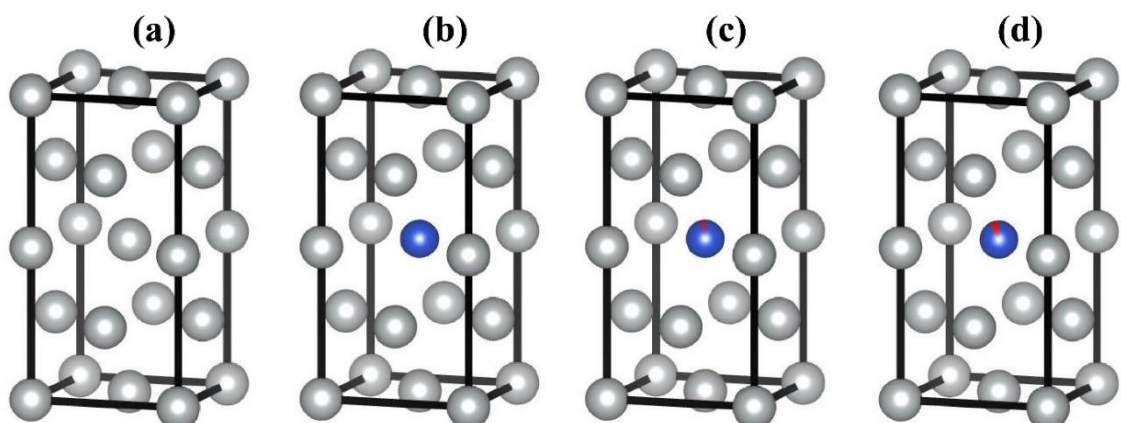

**Figure S3.** Crystal structure of (a) Ag, (b) ACZ0.0, (c) ACZ0.5, and (d) ACZ1.0.

**Table S2.** Results of Rietveld refinement for fine silver and ACZ0.5 alloy films

| Sample | Cell parameter Å | Volume Å³ | $R_p$ | $R_{wp}$ | $R_{exp}$ | GOF | $R_{Bragg}$ | $\chi^2$ | $R_f$ |
|---|---|---|---|---|---|---|---|---|---|
| FS | $a = b = c = 4.094$ | 68.619 | 12.1 | 8.0 | 13.25 | 1.1 | 1.02 | 1.14 | 1.54 |
| ACZ0.5 | $a = b = 4.074$, $c = 8.148$ | 135.236 | 12.3 | 7.9 | 11.23 | 1.0 | 1.13 | 1.03 | 1.29 |

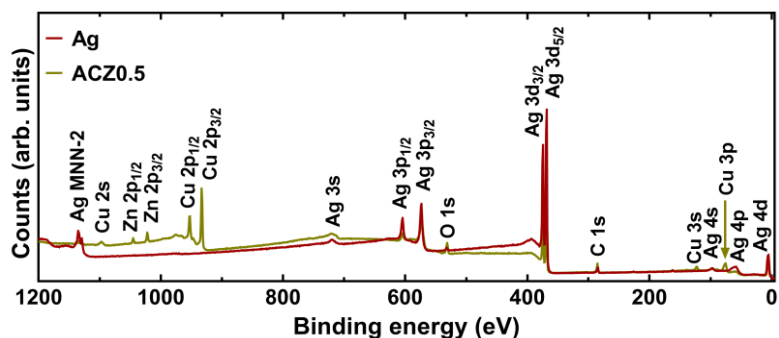

**Figure S4.** XPS full survey of Ag (control), and ACZ0.5 (target) electrode films.

The Raman spectra of Ag, ACZ0.0, ACZ0.5, and ACZ1.0 electrode films are recorded at room temperature and shown in **Figure S5**. The Ag Raman spectrum consists of vibrational mode at 239, 883, 1287, 1359, and 1606 $cm^{-1}$. In the Raman spectra of Ag, the band observed at 239 cm[-1], is due to $Ag-O$ stretching mode[4,5], the vibrational peak observed at 883 cm[-1] arises due to $Ag-O$ interaction through the hydrophilic part of carboxylic group[6,7]. The other band observed at 1287, 1359, and 1606 cm[-1] are arise due to carboxylic symmetric and anti-symmetric $C=O$ stretching vibration of carboxylic group, respectively[4,6]. The Raman spectrum of ACZ0.0 reveals a new mode at 617 $cm^{-1}$, associated with the bending of $C-O$ bonds[8,9], while the 239 $cm^{-1}$ mode of $Ag-O$ has vanish. The Raman spectra of ACZ0.5 and ACZ1.0 exhibit a new mode at 998 $cm^{-1}$, wherein the $C-O$ bending mode at 617 $cm^{-1}$ has ceased to be observed. This mode is associated with 2TO (transverse-optical) vibration mode of wurtzite ZnO[10,11], which confirms the formation of ZnO on the surface of thin films.

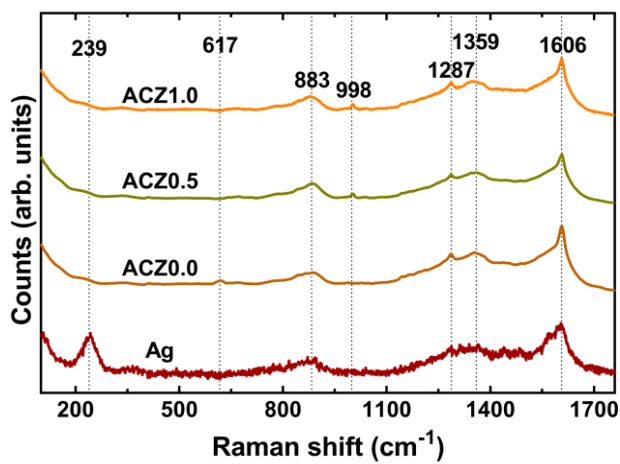

**Figure S5.** Raman spectra of Ag, ACZ0.0, ACZ0.5, and ACZ1.0 electrode films, which were obtained using a 532 nm excitation wavelength.

The UPS spectra of Ag and ACZ0.5 are shown in **Figure S6**. The determination of the energy associated with the secondary electron cutoff (SECO) entails an examination of the UPS spectrum when plotted in relation to the Fermi level. Assuming that Fermi level alignment is preserved between the spectrometer and the sample surface, the work function of the sample surface is given by[12]

$$\phi = h\nu - SECO \tag{1}$$

Equation (1) is certainly valid for a metal's surface or for an adsorbate on a conducting substrate. The work functions of Ag and ACZ0.5 are calculated to be 4.61 eV and 4.93 eV, respectively, as shown in **Figure S6a,b**.

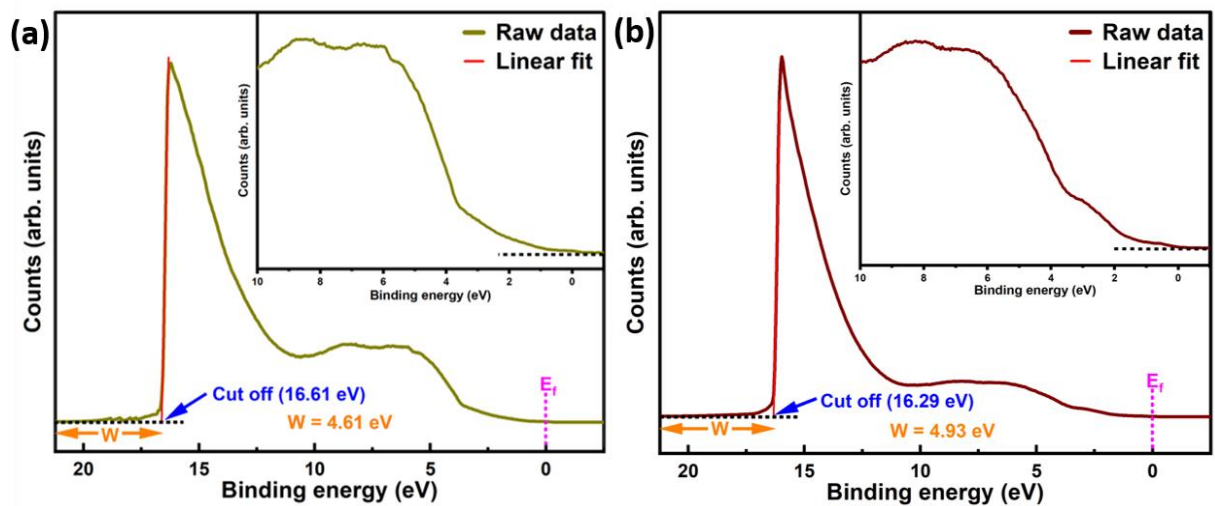

**Figure S6.** UPS spectra of (a) Ag (control) and (b) ACZ0.5 (target) films.

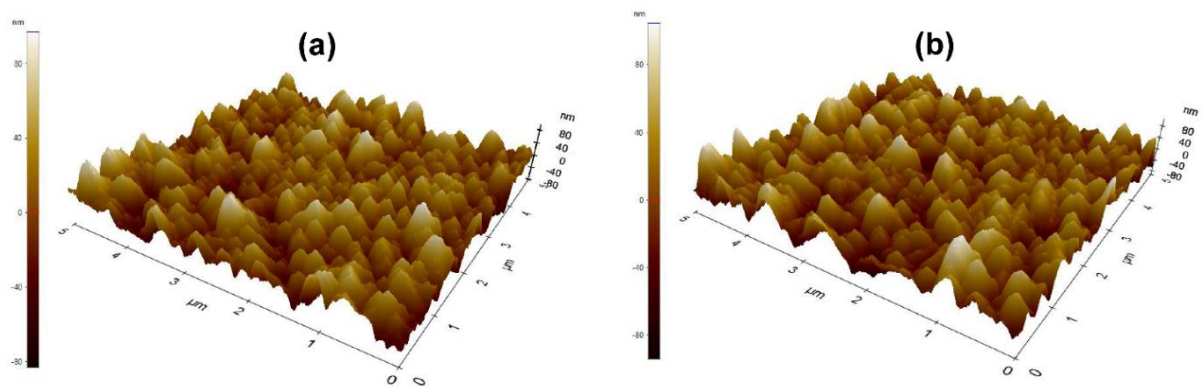

**Figure. S7** High-resolution 3D AFM images showing the surface topography of (a) Ag (control) and (b) ACZ0.5 (target) back electrode films.

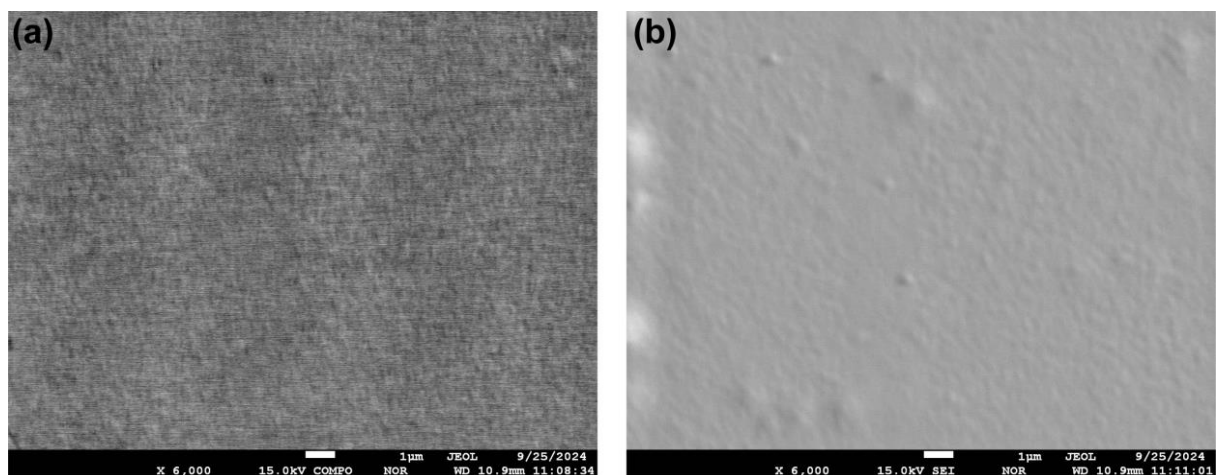

**Figure S8.** (a) BSE and (b) SE image of ACZ1.0 alloy thin film on glass substrate.

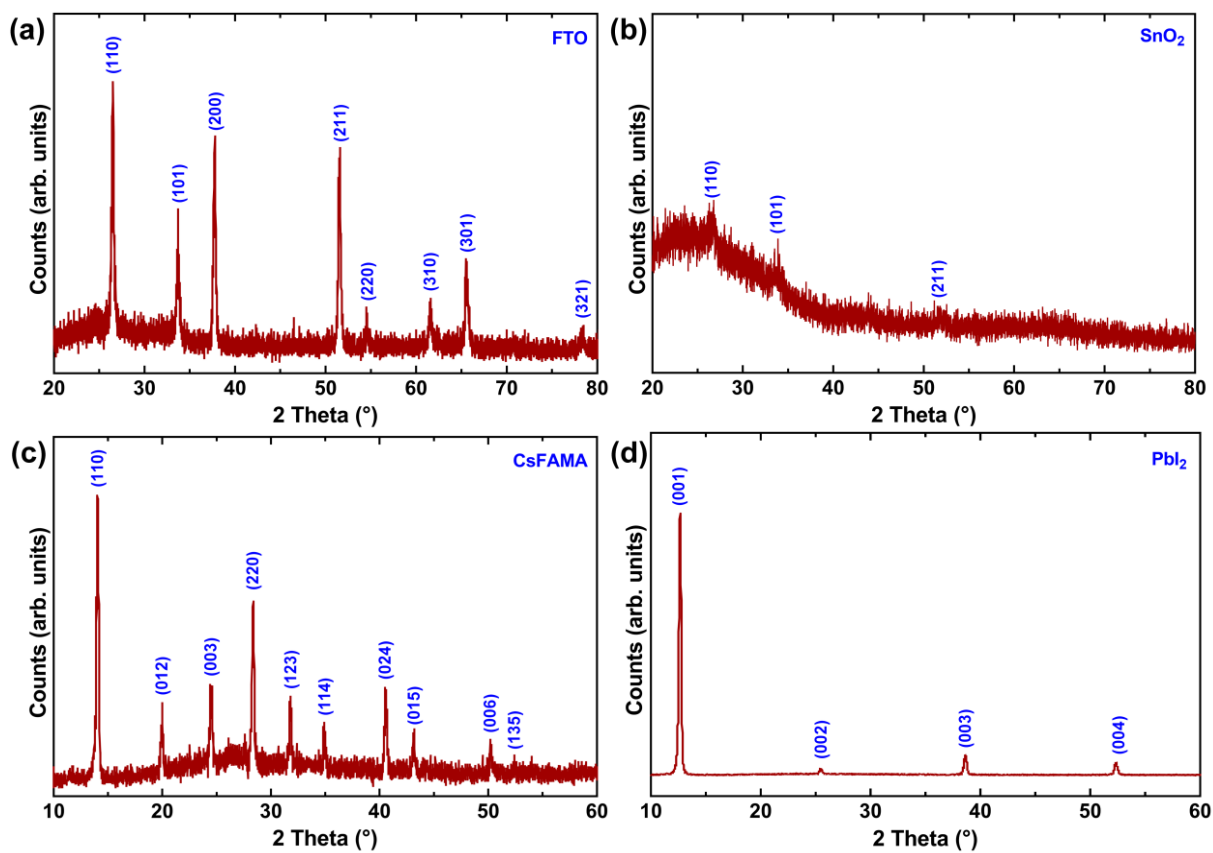

**Figure S9.** XRD patterns of (a) Fluorine doped $SnO_2$, (b) $SnO_2$, (c) CsFAMA, and (d) $PbI_2$ thin film on glass substrate.

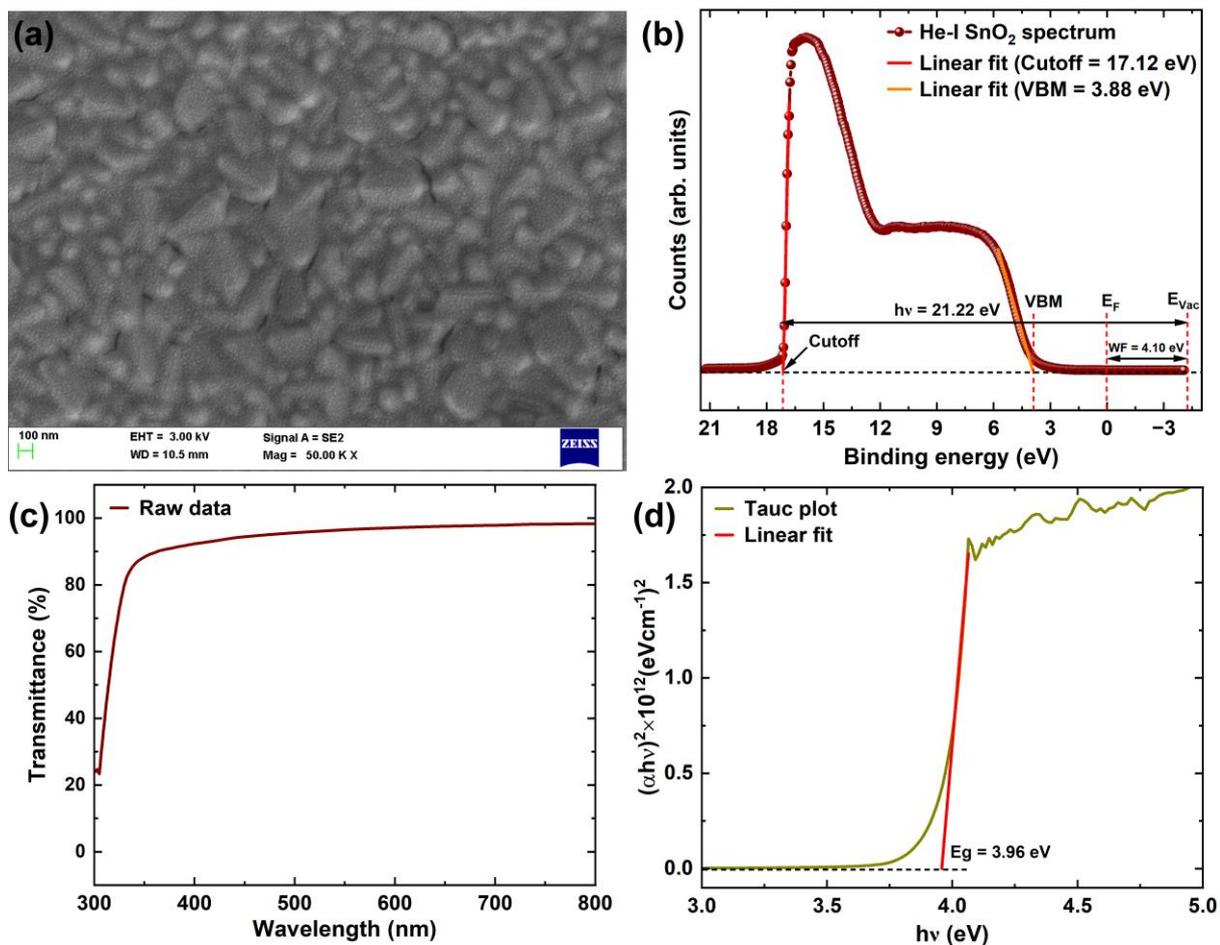

**Figure S10.** SEM image of SnO$_2$ thin film on FTO coated glass, (b) UPS spectra of SnO$_2$ thin film, (c) transmittance spectra of SnO$_2$ thin film, and (d) corresponding Tauc plot for band gap measurements.

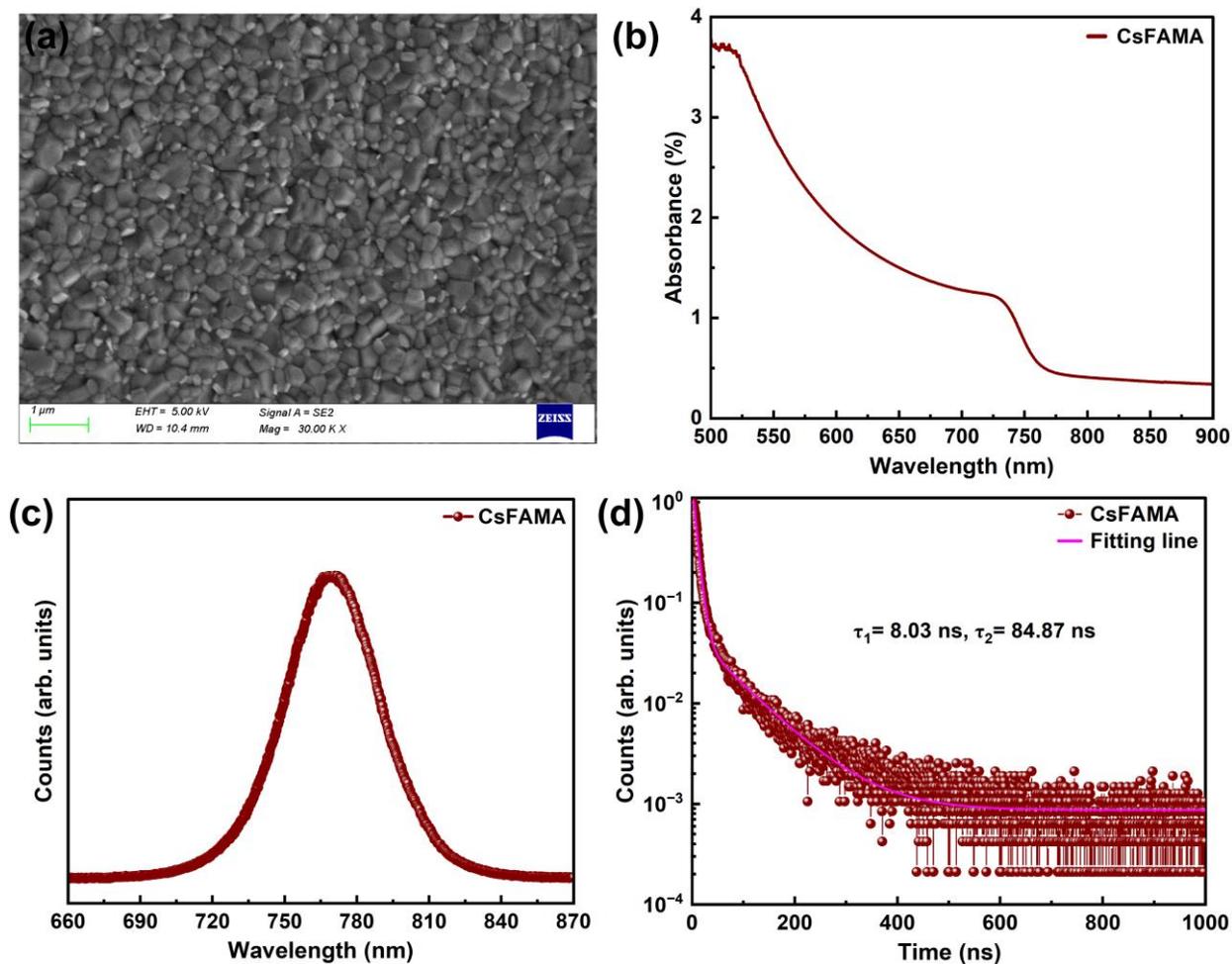

**Figure S11.** (a) SEM image (b) absorption spectra, (c) steady-state photoluminescence spectra, and (d) time-resolved photoluminescence profiles of CsFAMA thin film.

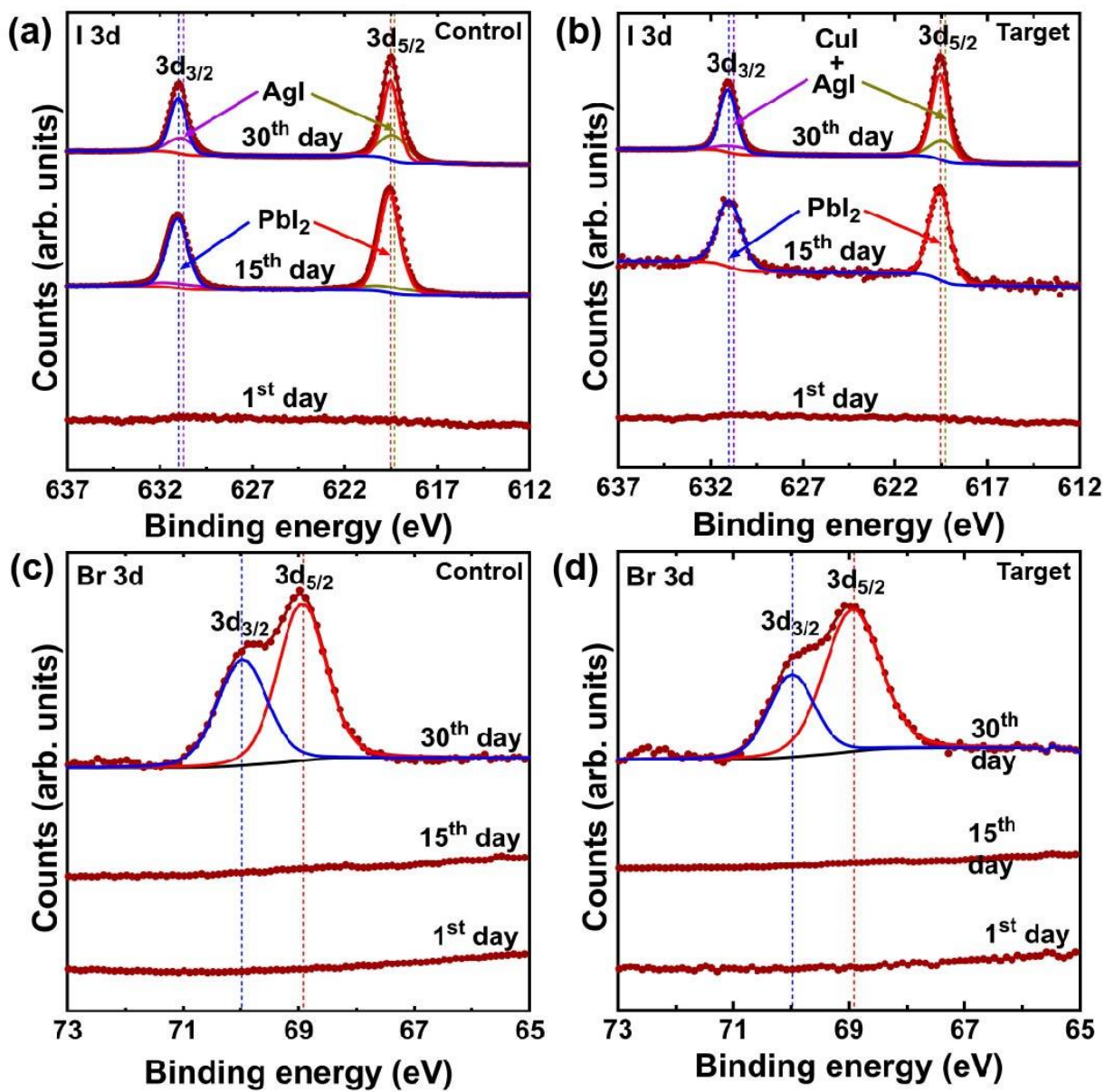

**Figure S12.** The I 3d XPS spectra of the device with (a) Ag (control), and (b) ACZ0.5 (target) electrodes, and the Br 3d XPS spectra of the device with (c) Ag (control), and (d) ACZ0.5 (target) electrodes.

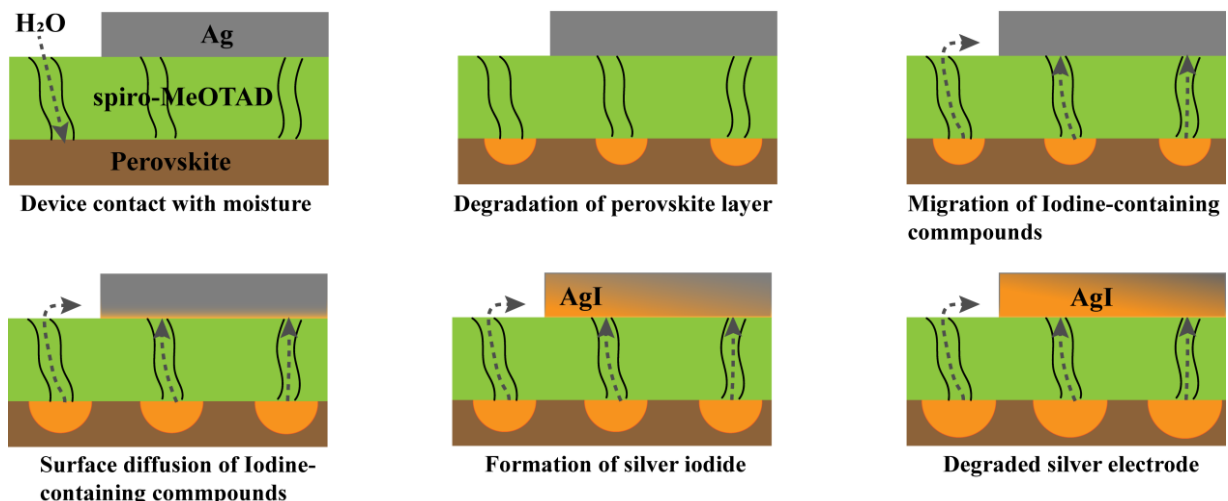

**Figure S13.** Conceptual model of degradation of Ag electrode deposited on the FTO/SnO$_2$/perovskite/spiro-OMeTAD based on XRD and XPS analysis.

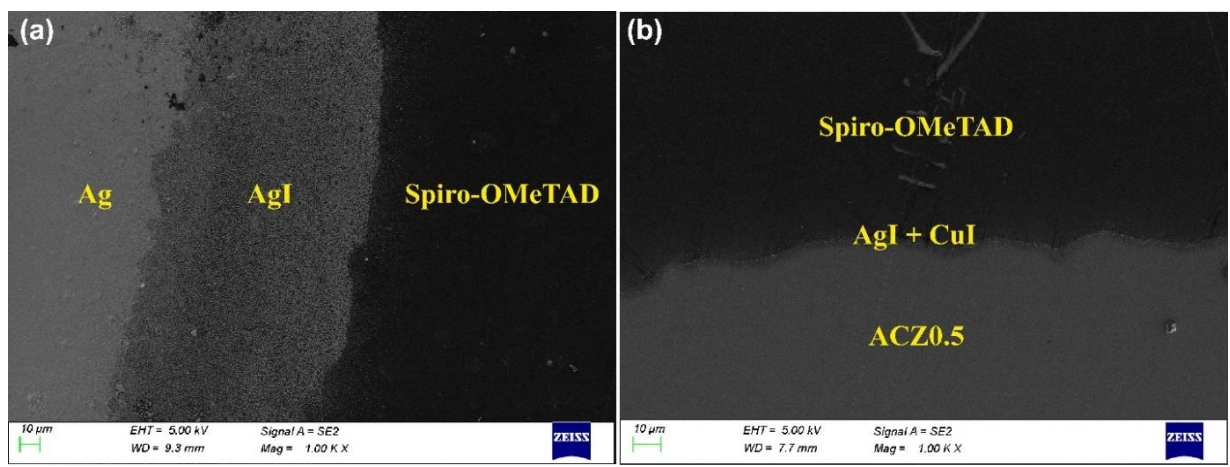

**Figure S14.** SEM images of the back electrodes of (a) the control device and (b) the target device after one month.

**Sheet resistance:** The sheet resistances of the silver and alloys films were calculated by the Van Der Pauw method. The values of the sheet resistance of Ag, ACZ0.0, ACZ0.5, and ACZ1.0 films were measured to be 0.012, 0.384, 0.425, and 1.144 ohm/sq, respectively.

## 3.2. Device characterization

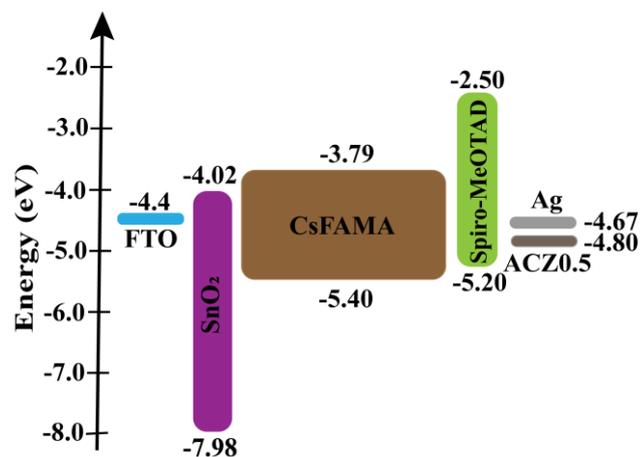

**Figure S15.** Energy band diagram of the PSCs with Ag and ACZ0.5 electrodes.

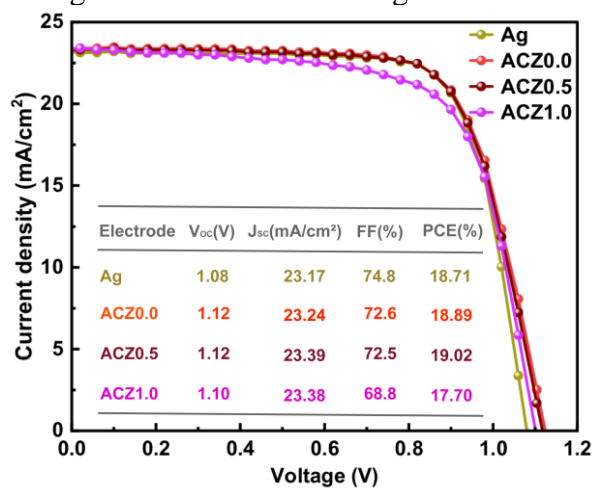

**Figure S16.** Illuminated current-voltage characteristics of PSCs of Ag, ACZ0.0, ACZ0.5, and ACZ1.0 electrodes with reverse scan.

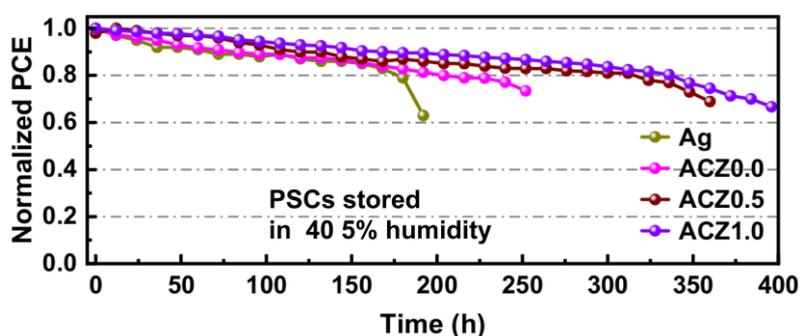

**Figure S17.** Stability measurements of the PSCs with Ag, ACZ0.0, ACZ0.5 and ACZ1.0 which are stored in an environment of 40 ± 5% humidity.

**Table S3.** *J–V* data of 28 independent devices.

| Control devices | Voc (V) | Jsc (mA.cm$^{-2}$) | FF (%) | PCE (%) | Target devices | Voc (V) | Jsc (mA.cm$^{-2}$) | FF (%) | PCE (%) |
|---|---|---|---|---|---|---|---|---|---|
| 1  | 1.08 | 23.17 | 74.8 | 18.71 | 1  | 1.12 | 23.39 | 72.5 | 19.02 |
| 2  | 1.08 | 23.23 | 73.0 | 18.34 | 2  | 1.12 | 23.45 | 71.8 | 18.87 |
| 3  | 1.09 | 23.11 | 71.9 | 18.15 | 3  | 1.12 | 23.23 | 72.0 | 18.76 |
| 4  | 1.08 | 22.89 | 72.9 | 18.07 | 4  | 1.13 | 23.67 | 69.7 | 18.66 |
| 5  | 1.08 | 22.57 | 73.5 | 17.95 | 5  | 1.12 | 23.34 | 71.1 | 18.62 |
| 6  | 1.07 | 22.34 | 74.6 | 17.87 | 6  | 1.11 | 23.05 | 72.5 | 18.56 |
| 7  | 1.07 | 22.63 | 72.7 | 17.65 | 7  | 1.12 | 22.89 | 71.8 | 18.43 |
| 8  | 1.08 | 22.46 | 72.2 | 17.56 | 8  | 1.11 | 23.17 | 71.4 | 18.37 |
| 9  | 1.07 | 22.18 | 73.8 | 17.54 | 9  | 1.11 | 22.46 | 73.5 | 18.33 |
| 10 | 1.07 | 22.86 | 71.2 | 17.45 | 10 | 1.10 | 22.57 | 73.3 | 18.24 |
| 11 | 1.07 | 22.47 | 72.1 | 17.36 | 11 | 1.10 | 22.32 | 73.9 | 18.17 |
| 12 | 1.06 | 22.34 | 72.7 | 17.24 | 12 | 1.10 | 22.87 | 72.0 | 18.13 |
| 13 | 1.07 | 22.07 | 72.8 | 17.23 | 13 | 1.11 | 23.15 | 70.2 | 18.06 |
| 14 | 1.06 | 21.84 | 74.0 | 17.16 | 14 | 1.10 | 22.57 | 72.5 | 18.02 |
| 15 | 1.06 | 21.94 | 73.1 | 17.03 | 15 | 1.10 | 22.64 | 72.1 | 17.98 |
| 16 | 1.05 | 22.23 | 72.2 | 16.89 | 16 | 1.10 | 22.03 | 73.7 | 17.87 |
| 17 | 1.05 | 22.09 | 72.5 | 16.86 | 17 | 1.09 | 23.34 | 69.7 | 17.74 |
| 18 | 1.06 | 22.27 | 70.9 | 16.76 | 18 | 1.10 | 21.93 | 73.1 | 17.65 |
| 19 | 1.05 | 21.67 | 73.4 | 16.74 | 19 | 1.09 | 22.42 | 71.9 | 17.58 |
| 20 | 1.05 | 21.48 | 73.8 | 16.67 | 20 | 1.10 | 21.78 | 73.2 | 17.56 |
| 21 | 1.06 | 21.67 | 71.5 | 16.45 | 21 | 1.09 | 22.23 | 72.0 | 17.45 |
| 22 | 1.05 | 21.36 | 72.3 | 16.24 | 22 | 1.10 | 21.69 | 72.7 | 17.36 |
| 23 | 1.05 | 21.21 | 72.5 | 16.17 | 23 | 1.09 | 23.34 | 67.2 | 17.11 |
| 24 | 1.04 | 21.57 | 71.1 | 15.97 | 24 | 1.09 | 22.04 | 71.0 | 17.06 |

| 25 | 1.05 | 21.08 | 70.2 | 15.57 | 25 | 1.10 | 22.56 | 67.4 | 16.75 |
| --- | --- | --- | --- | --- | --- | --- | --- | --- | --- |
| 26 | 1.05 | 21.46 | 68.5 | 15.45 | 26 | 1.08 | 23.05 | 66.9 | 16.67 |
| 27 | 1.04 | 20.82 | 69.4 | 15.06 | 27 | 1.08 | 22.38 | 68.0 | 16.45 |
| 28 | 1.04 | 21.04 | 66.4 | 14.56 | 28 | 1.09 | 21.89 | 65.9 | 15.76 |